# Temporal self-compression and self-frequency shift of sub-µJ pulses at 8 MHz repetition rate

Francesco Tani[1], Jacob Lampen[2], Martin Butryn[1], Michael H. Frosz[1],
Jie Jiang[2], Martin Fermann[2] and Philip St.J. Russell[1]

[1]*Max Planck Institute for the Science of Light, Staudtstr. 2, 91058 Erlangen, Germany,*
[2]*IMRA America, Inc., 1044 Woodridge Avenue, Ann Arbor, MI 48105, USA*
*francesco.tani@mpl.mpg.de*

**Abstract:** We combine soliton dynamics in gas-filled hollow-core photonic crystal fibers with a state-of-the-art fiber laser to realize a turn-key system producing few-fs pulses at 8 MHz repetition rate at pump energies as low as 220 nJ. Furthermore, by exploiting the soliton self-frequency shift in a second hydrogen-filled hollow-core fiber, we efficiently generate pulses as short as 22 fs, continuously tunable from 1100 nm to 1474 nm.

## 1. Introduction

Laser systems delivering trains of ultrashort pulses at MHz-level repetition rates and MW peak powers have become key to research and applications in numerous fields [1–3]. Recent advances in laser techniques and nonlinear optics have enabled efficient generation of pulses as short as a single cycle at average powers of 10s of W and energies from a few µJ up to the MJ level at repetition rates as high as 10 MHz [4–8]. Such remarkable specifications are typically achieved by temporally compressing ~100 fs pulses from solid-state lasers by means of hollow-core photonic crystal fibers (HC-PCFs) and capillaries, multi-pass cells, and glass plates [9]. Reaching durations of a few fs requires, however, large compression ratios (~100) and multiple stages, which is not ideal as it results in increased system complexity and reduced throughput efficiency. Furthermore, achieving good stability is challenging at very high repetition rates, when the increased average power can cause thermal instabilities. Few-fs pulses with lower energies (a few tens of nJ up to a few hundred nJ) are typically sufficient for studying and controlling ultrafast dynamics in condensed matter systems [1–3]. Nevertheless, such pulses cannot be delivered directly by laser oscillators, and post-compression using solid-core fibers is highly inefficient and complex, requiring compensation of higher-order phase terms to reach the few-fs regime [10–12]. Moreover, most systems delivering amplified pulses with durations of a few 10s of fs are centered at 0.8 µm, 1 µm, and 2 µm wavelengths, and lack the spectral tunability that is often achieved via optical parametric amplification.

Gas-filled HC-PCF provides a versatile platform both for generating few-fs pulses by temporal compression and for efficient frequency conversion [13,14]. Recent work has reported exciting new opportunities for using stimulated Raman scattering to efficiently generate ultrashort pulses that are wavelength-tunable in the near-infrared (IR) [15–18]. So far, however, in hollow capillaries the large core and high loss have constrained Raman-driven dynamics to the normal dispersion regime. On the other hand, anomalous dispersion can be easily achieved in HC-PCF, permitting access to soliton dynamics such as the Raman-driven soliton self-frequency shift (SSFS), [15,16,18], which have long been studied and exploited in solid-core fibers. In comparison, SSFS in HC-PCF allows shorter-wavelength pulses to be redshifted by an amount that can be adjusted by varying the gas pressure. Furthermore, it is possible to achieve considerably shorter pulse durations and higher pulse energies than in solid-core fibers, where the minimum pulse durations have been limited to ~80 fs, and the maximum energies to ~90 nJ [19–21].

Here we report efficient generation of 5 fs pulses (1030 nm, 8 MHz) by launching 40 fs pulses with energies in the range 220 nJ to 475 nJ, delivered from a state-of the-art fiber laser, into a gas-filled HC-PCF. We then demonstrate, in a second experiment with a $H_2$-filled HC-PCF, that the pulse wavelength can be efficiently and continuously tuned across the near IR by the SSFS. In this fashion, we generated pulses as short as 22 fs pulses at energies between 120 nJ and 160 nJ (~0.95 W to 1.28 W average power) at wavelengths from 1080 nm to 1474 nm. Finally, for the range of parameters here considered, both processes are highly coherent, so that carrier-envelope phase (CEP) stability could be conveniently achieved by locking the carrier-envelope offset (CEO) frequency $f_{CEO}$.

## 2. Laser system & soliton-effect self-compression

The experimental setup is shown in Fig.1. The laser is a highly coherent 80 MHz Yb-fiber oscillator mode-locked at near-zero dispersion using a nonlinear amplifying loop mirror [22]. The system includes a down-counter to 8 MHz, which permits a CEP-stable output when the $f_{CEO}$ is locked at 16 MHz. The pulse train is then amplified in a large mode area, polarization-maintaining Yb-doped fiber (mode area 500 µm$^2$, length 3.0 m) by self-similar amplification [23]. In this fashion, we obtain 1030 nm ~40 fs full width half maximum (FWHM) pulses with average powers up to 7.8 W (~0.97 µJ pulse energy). A wire grid polarizer is used to control the output power, followed by a half-wave plate to match the fiber polarization axis and a plano-convex lens to couple into the HC-PCF. Scanning electron micrographs (SEMs) of the two HC-PCFs used in the experiments are shown in Fig. 1(b). The PCF on the right was employed for pulse compression (denoted as fiber A in what follows) and the second one for SSFS (denoted as fiber B in what follows). In both experiments, the HC-PCFs were placed in gas-cells with a coated fused silica window at the input and uncoated windows at the output (a 1 mm thick $MgF_2$ window after PCF A, and a 5 mm thick fused silica window after PCF B). The light emerging from the fibers was collimated using a silver parabolic mirror, after which the output spectra were recorded using an integrating sphere connected to a spectrum analyzer (Yokogawa) via a multimode fiber. For temporal characterization of the pulses, we used second-harmonic-generation frequency-resolved optical gating (SHG FROG) in an all-reflective non-collinear geometry, with a 10 µm thick BBO crystal as the nonlinear medium.

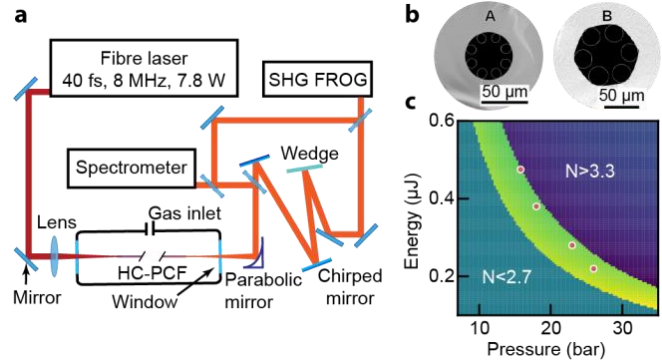

Figure 1. (a) Sketch of the experimental setup, b) scanning electron micrographs of the two single-ring HC-PCFs. The fiber on the right (fiber A) has a core diameter of 30 µm, a core-wall thickness of ~220 nm and was used for temporal compression. The fiber on the left (fiber B) has a core diameter of 40 µm and a core-wall thickness of ~320 nm and was used for SSFS. c) Calculated compression length as a function of pulse energy and argon-filling pressure. The calculation is for fiber *a* filled with argon and soliton order $2.7 < N < 3.3$, the light green/yellow area marks the range of energies and pressures for which $L_C = 18 \pm 1.5$ cm and the red dots denote the experimental measurements.

Both HC-PCFs have anomalous dispersion ($\beta_2(\omega_0) < 0$) at the laser frequency for Ar (and $H_2$) filling pressures up to several tens of bar. As a result, soliton dynamics can be conveniently accessed by launching laser pulses with peak power $P_0$ and duration $T_0$ such that the soliton order $N = \sqrt{\gamma P_0 T_0^2/|\beta_2|} \geq 1$, where $\gamma$ is the fiber nonlinear coefficient [24] and $\beta_2$ depends on the species and pressure of the gas. During propagation higher-order solitons undergo self-compression, their duration shortening by a factor approximately proportional to $N$ over the length $L_c = (1/2N + 1.7/N^2)L_D$, where $L_D$ is the dispersion length [24]. Beyond this point, the input pulses undergo either soliton fission or a periodic evolution depending on the perturbations introduced by effects such as higher-order dispersion, self-steepening and ionization. The distance $L_C$ over which the pulses remain temporally focused (typically by a few cm) can be adjusted by varying the pressure of the filling gas, the energy of the input pulses, or adding a positive chirp. As a result, a fiber of fixed length can be employed to obtain the same compression factors over a broad range of input energies. In Fig. 1(d), we show the range of pulse energies and Ar-pressures for which ~8-fold compression can be achieved in an 18-cm-length of fiber A, the red dots indicating the parameters used in the experiment. Because the full-width half-maximum (FWHM) of the pump laser pulses is already only a few tens of fs, we can obtain single-cycle durations at low soliton orders ($N \sim 3$), thus ensuring efficient and stable temporal compression. The plot in Fig. 1(d) shows that by changing the gas pressure, the same fiber sample can be used to compress 40 fs pulses with energies between ~600 nJ and ~150 nJ, and that much wider energy ranges can be readily accessed using HC-PCFs with different core radii and lengths.

By exploiting the scale invariance of soliton dynamics in gas-filled HC fibers [5], self-compression to a few fs has been demonstrated over a wide range of pulse energies from a few µJ up to the mJ-level [4,5,25–27]. Here, we show that the same dynamics can be reproduced using input pulses with energies as low as ~200 nJ. To demonstrate this, we used an 18 cm length of fiber A in which the capillaries surrounding the core had wall thickness ~220 nm, which was thin enough to avoid wavelength-dependent distortions to the dispersion profile (caused by anti-crossings between wall resonances and the core mode) and ensure high fidelity temporal compression [28]. It is worth noting that when HC fibers are pumped by MHz pulse trains with W-level average power, thermal effects in the gas can affect the pulse-to-pulse stability [4,29,30]. These instabilities are caused by nonlinear absorption and ionization, which are enhanced by contamination of the filling gas by, for example, water. They can be suppressed to a large extent by evacuating and purging the fiber with an inert gas before the experiment and selecting the experimental parameters to minimize photoionization.

A plano-convex lens with a 10 cm focal length was used to launch the laser pulses into the HC-PCF (Fig. 1(a?)), resulting an overall transmission of ~74%, in part limited by the distorted beam profile (typically >90% can be achieved in the absence of beam distortion).

Even though the few-cycle pulses at the fiber output are transform-limited, dispersion compensation is required to measure them because propagation through the gas in the cell, its output window, and in air afterwards introduces a positive chirp, which was compensated using a pair of double-angle negatively chirped mirrors (UltraFast Innovations GmbH, group delay dispersion of −80 fs$^2$ per pair) in combination with thin fused silica wedges (Altechna) for dispersion fine-tuning. Upon filling the HC-PCF with Ar and tuning the pressure over the range 15.8 to 26 bar, we could obtain similar soliton orders ($N$~3) for input pulses with energies between 220 nJ and 475 nJ and thus achieve comparable pulse durations at the fiber output. This is shown in Fig. 2(a) & (b) where we plot the retrieved temporal and spectral profiles of the compressed pulses at pump energies of 220 nJ, 280 nJ, 380 nJ and 475 nJ. The FWHM duration is between 5.2 fs and 5.8 fs, with more than 76% of the energy within the main peak. The measured and retrieved FROG traces obtained after filling the fiber with 23 bar of Ar are shown in Figs. 2(c) & (d).

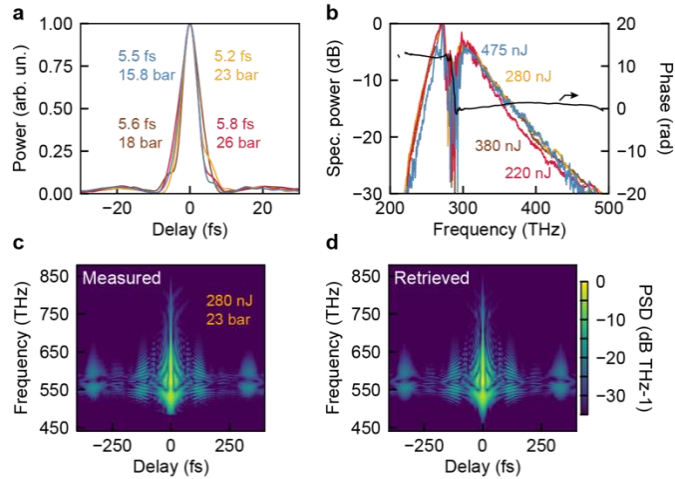

*Figure 2* (a) Temporal and (b) spectral profiles of the pulses at the output of an 18 cm length of fiber A. The curves were obtained from SHG-FROG measurements for a selection of input pulse energies and H$_2$ filling pressures. The bottom panels show the (c) measured and (d) retrieved FROG traces when 280 nJ pulses are launched into the fiber filled with 23 bar of H$_2$.

## 3. Soliton self-frequency shift in H$_2$-filled SR-PCF

To demonstrate a SSFS-based source of ~30 fs pulses with tunable central wavelength, we used 1.6 m and 2.9 m lengths of fiber B filled with H$_2$. The hollow-core diameter was 40 μm, the core-wall thickness lay between ~275 nm and ~320 nm, and the fiber was coiled up in one turn with a ~24 cm radius. Compared to fiber A, the core-wall thickness was thicker in order to shift the first low-loss transmission window (located at $\lambda > \lambda_{ac} = 2t\sqrt{n_{silica}^2 - n_{core}^2}$ where $t$ is the core wall thickness, $n_{silica}$ the index of silica, and $n_{core}$ the index of the filling gas [31,32]) to longer wavelengths so as to accommodate the red-shifting soliton. Using a plano-convex lens with 12 cm focal length, we launched 500 nJ (corresponding to 4 W) pulses into the HC-PCF and achieved ~65% transmission, which is lower than the 74% obtained with fiber A, mainly because of a slightly lower coupling efficiency and higher overall loss due to the much longer fiber length (the loss of the PCF at the pump wavelength, measured by cut-back, was <0.25 dB/m). Using a half-wave plate to align the linearly polarized pump field with one of the birefringent eigenaxes of the HC-PCF. This was crucial for preserving linear polarization state, which can be impaired by the long fiber length and the rotational Raman nonlinearity, which can be reduced but not entirely suppressed when the pulses are shorter than the period of rotational Raman scattering (56 fs). After adjustment, the light exiting the fiber was >99% linearly polarized.

To illustrate the SSFS in H$_2$-filled single-ring HC-PCFs, Fig. 3(a)&(b) shows the simulated temporal and spectral pulse evolution in fiber B filled with 40 bar H$_2$. A unidirectional nonlinear wave equation [33] with the Raman polarization term represented by a response function was used in the modeling, as discussed in [34]. The optical loss was included using the empirical formula given in [32], and as input we used the pulse parameters measured by the SHG-FROG at an energy of 220 nJ (soliton order ~2.6). Although under these conditions photoionization of the gas is almost negligible, we nevertheless included it using the Perelomov-Popov-Terent'ev (PPT) rate, modified with the Ammosov-Delone-Krainov (ADK) coefficients [35].

For a fixed set of gas and laser parameters, the red-shift increases with fiber length, but ultimately it is limited by the waveguide loss and absorption in $H_2$. The pump pulse first undergoes soliton-effect self-compression, and then after about 40 cm of propagation, a red-shifting soliton emerges, shifting to longer wavelength, reaching 1385 nm after 1.6 m and 1480 nm after 2.9 m. At this last position, ~18.7% of the input pulse energy has been absorbed by the $H_2$ and the red-shifted soliton carries about 53% of the output power. Besides fiber loss and the $H_2$ absorption, the fraction of the input energy carried by the red-shifted component is affected by the soliton order $N$, higher values resulting in a lower red-shifted fraction. According to simulations, this fraction varies between 87% and 53% when the pressure is varied from 5 bar to 40 bar, corresponding to soliton orders between ~1.0 and ~3.7. The simulated and measured spectra are in excellent agreement. In Fig. 3(c) we show the spectra measured at the output of the fiber filled with $H_2$ in the same pressure range, and compare the measured and simulated spectra in PCFs of length 1.6 m (Fig. 3(d)) and 2.9 m (Fig. 3(e)), filled with 40 bar $H_2$. For the selected experimental parameters, the wavelength can be continuously tuned between 1080 nm and 1474 nm, while the red-shifted solitons carry over 0.95 W of average power (~120 nJ) and up to 1.28 W (~160 nJ) at 1300 nm. We also found that a lower pulse energy (by ~30%) was needed in the simulations to reproduce the measured spectra, in agreement with the experimental observations that, compared to theory, a larger fraction of energy is lost during propagation, resulting in a slightly lower conversion efficiency to the red-shifted peak. We attribute this to excitation of higher order modes at the fiber input, which causes the appearance of a tall and narrow spectral peak around 1030 nm that is not reproduced in the numerical simulations. Using a 10 nm wide bandpass filter to isolate this spectral region we were able to confirm that at this wavelength the light was indeed in a higher-order mode (HOM) and carried about 15% of the output power. Notwithstanding, the radiation in the rest of the spectrum was in a pure fundamental mode. This observation may well explain the discrepancy between simulations and experiments since the optical loss is much higher for HOMs.

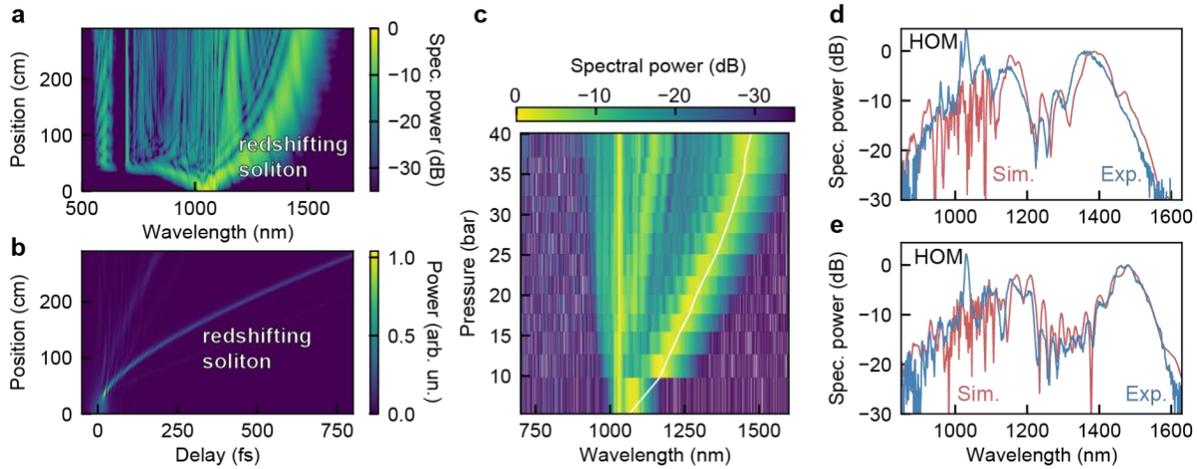

Figure 3 Simulated (a) spectral and (b) temporal evolution of a 220 nJ pulse in fiber B filled with 40 bar of $H_2$. (c) Measured spectra at the fiber output as a function of the $H_2$-filling pressure with the white line indicating the average wavelength of the red-shifted soliton. (d) Measured (blue) and simulated (red) spectra for a 1.6 m length of fiber filled with 40 bar of $H_2$. (e) Same as (d) but using a 2.9 m long fiber. The spectra are normalized to the peak of the red-shifted spectral components. As input pulses for the numerical simulations, we use the measured temporal profile with an energy of 220 nJ.

This LP02-like HOM does not undergo any spectral broadening nor does it affect the SSFS. Due to its different group velocity, in fact, it walks off quickly from the main pulse. Moreover, the larger negative dispersion of the HOM results in $N < 1$, thus preventing any soliton dynamics. We attribute excitation of the HOM to transverse variations in refractive index caused by absorption of light in $H_2$ [30]. The narrow bandwidth of the HOM signal suggests that it is excited close to the fiber input; if it were generated further along the fiber by intermodal four-wave mixing, a larger bandwidth would be expected since the pump pulse will have undergone spectral broadening. In contrast with the Ar-filled fiber, when no HOM was observed at the output, a HOM appeared every time in the $H_2$-filled PCF in a narrow spectral region around 1030 nm, despite careful alignment. Measurements of the laser power after the fiber (Fig. 4(a)) further support the hypothesis. The total output and the red-spectral component power both decrease steadily as the $H_2$ filling pressure goes up. The waveguide loss, which is higher at longer wavelength, gas absorption, and the higher soliton order (which increases with pressure) only partially explain this behavior. At the same time the degree of polarization degrades at higher pressure, and at 40 bar ~10% of the output power was polarized orthogonally to the red-shifted soliton, which was still linearly polarized and in the fundamental mode. In Fig. 4(b), we show the average wavelength and power of the red-shifted spectral component after lengths of 1.6 m and 2.9 m. The power was obtained from calibrated spectral measurements and further confirmed by using long-pass filters followed by a thermal power meter. In the shorter PCF, the power of the red spectral component at a selected wavelength was up to 10% higher, but the maximum redshift was limited to 1370 nm (1474 nm for the longer fiber). Like in solid-core fibers, where the red-shift is proportional to the nonlinear coefficient $\gamma$ [24], here, the red-shift is approximately linear with gas pressure (like $\gamma$). Except for an offset due to optical loss, for both fiber lengths the power in the red-shifted component is approximately inversely proportional to the soliton order $N$. This is because the red-shifting soliton emerges after the pulse has undergone soliton-effect self-compression, and the fraction of power carried by the shortened pulse is determined by the quality factor which goes like $\sim 1/N$.

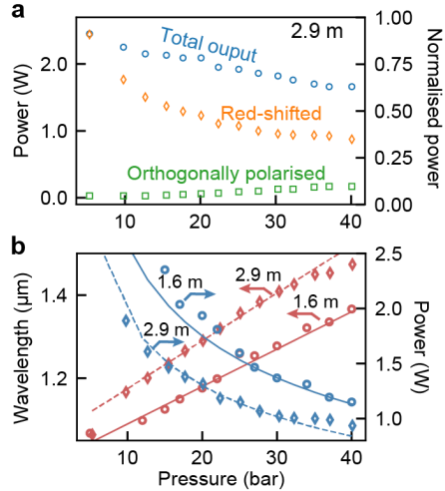

Figure 4 (a) Power recorded as a function of the H$_2$-filling pressure in a 2.9 m length of fiber B. The blue dots mark the total output power, the orange diamonds the power carried by the red-shifted spectral component and the green squares the output power of light polarized orthogonally to the input polarization axis. The right-hand axis indicates the power normalized to the total output power when the fiber is filled with 1 bar of H$_2$. Panel (b) shows the central wavelength (red symbols) and the power (blue symbols) of the red-shifted solitons in 1.6 m (circles) and 2.9 m (diamonds) lengths of fiber B. In both panels the full curves are meant as guides to the eye. The blue lines denote fits of the data to $A/N$, where $A = 3.5$ for the solid line and $A = 2.6$ for the dashed line. The red lines are linear fits of the wavelengths as a function of pressure.

An SHG-FROG was used for temporal characterization of the output pulses, placed directly after the parabolic mirror used for recollimation and an achromatic half-wave plate. The dispersion introduced by the optical components (between –20 and +10 fs$^2$/mm in the spectral region 1100 to 1500 nm, affecting the pulse duration by only a few fs) was accounted for in the data post-processing. The retrieved FROG traces reveal red-shifted pulses with a clean Gaussian-like temporal profile (Fig. 5(a)) and FWHM durations as short as 22 fs. These pulses arrive a few 100s of fs after the higher frequencies, a delay that increases with gas pressure and fiber length. Figure 5(b) shows the measured FWHM duration of the red-shifted pulses as a function of pressure for the two fiber lengths. In the 1.6-m-long PCF the red-shifted pulses shorten with increasing H$_2$ pressure, stabilizing at 24 ± 1.5 fs above 20 bar. For the 2.9-m-long PCF, on the other hand, the red-shifted pulse duration reaches a minimum (~27 fs) when filled with 15 bar H$_2$, slowly increasing for higher pressures and reaching 34 fs at 40 bar. By reducing the power and the frequency down-shift, optical loss also affects the soliton dynamics and thus the duration of the red-shifted pulses. Because the pulse energy falls with propagation along the fiber, $N$ eventually becomes less than 1, when the red-shifting pulse stops being a soliton and acquires a chirp.

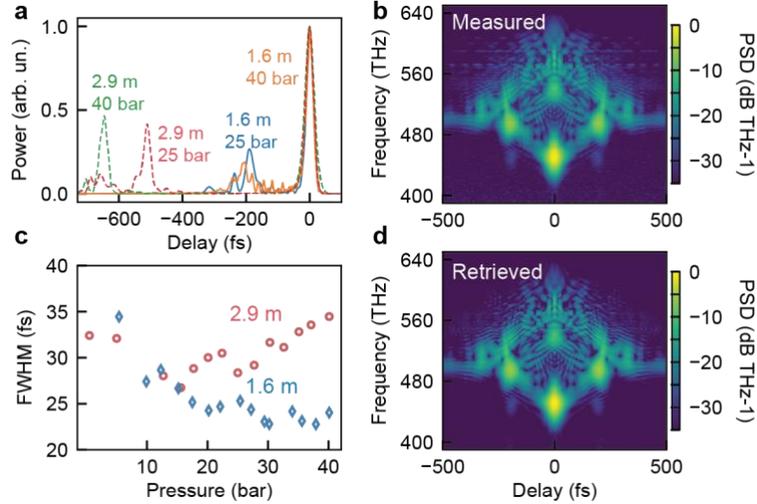

Figure 5 (a) Measured temporal pulse profile after the 1.6-m-long PCF (solid lines) and the 2.9-m-long PCF (dashed lines) at 25 bar and 40 bar of $H_2$. (c) Measured FWHM pulse duration at the output of the two fibers as a function of pressure. (b) Measured and (d) retrieved FROG traces at the output of the 1.6-m-long PCF filled with 30 bar of $H_2$.

In conclusion, soliton self-compression in gas-filled HC-PCFs, combined with a state-of-the-art fiber laser delivering 40 fs pulses with energies in the few 100s of nJ range and at a repetition rate of 8 MHz, results in a simple, efficient, and stable source of 5 fs pulses. By exploiting the soliton self-frequency shift in hydrogen-filled HC-PCF, pulses as short as 22 fs with energies in the range 120 to 160 nJ (average power up to 1.28 W) can be generated at wavelengths continuously tunable between ~1100 and ~1500 nm. This spectral region is important for multiphoton imaging of biological samples. These pulses have much higher energy and several times shorter durations than previously achieved using SSFS in waveguides [18–21]. The conversion efficiency to the red-shifted pulses varies between ~30% and 50% and is limited by the optical loss (mostly absorption by $H_2$), in-coupling losses and excitation of higher-order mode through thermal effects. The efficiency can be further increased by optimizing the launch efficiency (currently >70%) and using a pressure gradient to lower the pressure at the fiber input, thus mitigating thermal effects in the gas and further increasing the energy in the red-shifted solitons.